\documentstyle[12pt]{article}
\newcommand{\Eq}[1]{Eq.(\ref{#1})}

\newcommand{\refc}[1]{Ref.~\cite{#1}}
\newcommand{\refs}[1]{Refs.~\cite{#1}}

\newcommand{\bea}{\begin{eqnarray}}
\newcommand{\eea}{\end{eqnarray}}
\newcommand{\be}{\begin{equation}}
\newcommand{\ee}{\end{equation}}
\newcommand{\bc}{\begin{center}}
\newcommand{\ec}{\end{center}}
\newcommand{\ba}{\begin{array}}
\newcommand{\ea}{\end{array}}
\newcommand{\btab}{\begin{tabular}}
\newcommand{\etab}{\end{tabular}}
\newcommand{\bfig}{\begin{figure}}
\newcommand{\efig}{\end{figure}}

\newcommand{\BOX}{\hbox {$\sqcap$ \kern -1em $\sqcup$}}

\def\bigid{\leavevmode\hbox{\small1\kern-3.8pt\normalsize1}}
\def\id{\leavevmode\hbox{\small1\kern-3.3pt\normalsize1}}

\newcommand{\cL}{{\cal L}}
\newcommand{\tr}{{\rm tr\,}}
\newcommand{\fort}[3]{{\it Fortsch. Phys. }{ {\bf #1}{(#2)}{#3}}}

\newcommand{\mpl}[3]{{\it  Mod. Phys. Lett. }{{\bf #1} {(#2)} {#3}}}

\newcommand{\np}[3]{{\it  Nucl. Phys. }{{\bf #1} {(#2)} {#3}}}
\newcommand{\pr}[3]{{\it Phys. Rev.}{{ \bf #1} {(#2)} {#3}}}

\newcommand{\prl}[3]{ {\it Phys. Rev. Lett.}{{ \bf #1} {(#2)} {#3}}}

\newcommand{\prep}[3]{{\it Phys. Rep. }{{\bf #1} {(#2)} {#3}}}

\newcommand{\lm}{\tilde\cL^{\rm eff}(B,\mu)}

\normalsize
\begin{document}
% ==================================================
\large
\thispagestyle{empty}
\begin{flushright}   FIAN/TD/94-01 \\
		 G\"oteborg ITP 94--11\\
                    March 1994 \vspace{3ex} 
\end{flushright}
\bc
\normalsize
{\LARGE\bf A Note on  QED with  Magnetic Field and Chemical Potential\\}
\vspace{3ex}
{\Large David Persson\footnote{E-mail address: tfedp@fy.chalmers.se} }\\
{\large Institute of Theoretical Physics, 
  Chalmers University of Technology \\ and  G\"oteborg University ,  
 S-412 96  G\"oteborg} \vspace{2ex}\\
{\Large Vadim Zeitlin\footnote{E-mail address: zeitlin@td.fian.free.net}}\\
{\large Department of Theoretical Physics, P.~N.~Lebedev Physical Institute, \\
  Leninsky prospect 53, 117924 Moscow, Russia}\vspace{5ex}  
\ec
\centerline{{\Large\bf Abstract}}
\xiipt
\begin{quote}
Using a  generalized proper-time method,  we obtain expressions for the 
fermion density and the  QED effective 
Lagrangian for an  external magnetic field at finite  chemical potential.
The  effective  Lagrangian and the density are here written 
in terms of elementary functions,   summed  over a finite number of 
filled Landau levels.
\end{quote}
\newpage
% -----------------------------------------------------------------
\normalsize \setcounter{page}{1}
The study of finite temperature and density quantum electrodynamics
 (QED) 
with a nonvanishing average magnetic
field is of considerable interest as it may be associated with 
the electron-positron plasma
in  compact stellar objects (e.g. neutron stars and  magnetic white dwarfs),
where the fermion density and the magnitude of the magnetic field may 
be extremely high (see e.g. \refs{Chanm92,ShapiroT83}).

The  QED thermodynamical potential at finite temperature and 
density with a static  uniform magnetic field was  
   calculated already 25 years ago in \refc{Canuto68},  and using a
 generalization of 
Fock--Schwinger's proper-time method
for $ T, \mu \ne 0$  (where $\mu$ is the chemical potential) later in 
\refc{Cabo81}.
The interest to this problem was
 renewed 
in Ref.~\cite{ChodosEO90}, where an elegant
generalization of Fock--Schwinger's proper-time method in
 the case of nonzero magnetic
field and chemical potential was made. Using a real--time thermal formalism,
the results of \refc{ChodosEO90} were 
completed and generalized  to
finite temperature in \refc{ElmforsPS93}.
The expressions obtained for the effective action in the above cited
 references were
rather complicated and did include  some proper-time like integrals and/or 
infinite sums.

 Here we want to demonstrate that using the formalism,
elaborated in \refc{ChodosEO90} we may move forward and for the finite density
 QED with an  external
magnetic field at zero temperature show that it is possible to obtain
 simple expressions for the 
fermion density and the  effective Lagrangian. 
The effective Lagrangian is here written in terms of elementary functions 
 as a sum  over a  finite number of
(partially) filled Landau levels, and agrees with the zero temperature limit 
of the fermion partition function. As an application we confirm 
 that the magnetisation obtained from this
effective action does exhibit the relativistic de Haas--van Alphen effect
\cite{Canuto68,ElmforsPS93,LeeCCC69}.
\bigskip

We shall here consider finite density  QED with a nonvanishing 
average magnetic field. Including the chemical potential 
the corresponding Lagrangian reads
\be
	\cL = -\frac14 F_{\mu\nu}F^{\mu\nu} +\bar{\psi}(i {\partial
         \kern-0.5em/} -e {A\kern-0.5em/}  - \gamma_0 \mu -m)\psi~~~,
\label{eq-lagrange}
\ee
where we have chosen the  gauge $A_\mu = (0, -x_2B, 0, 0)$.

In order to calculate the one-loop correction to the effective Lagrangian,
$\int d^4\! x \cL^{{\rm eff}}= $\\ 
	$-i \ln \, {\rm Det} ( i{\partial \kern-0.5em/} -
	e{A\kern-0.5em/} - \gamma_0\mu - m)$,
we shall first evaluate the fermion  density 
	$\rho = \frac{\partial \cL^{{\rm eff}}}{\partial \mu}$.
We may  then reconstruct the effective Lagrangian according to 
\be
  \cL^{{\rm eff}} (B, \mu) =   \cL^{\rm eff}(B)+  \lm~~~,
\ee
where
\be
    \lm = \int_0^\mu \rho (B, \mu' ) d\!\mu' ~~~,
\ee
is  the contribution due to the finite density, and\vspace{1ex}
\be
  \cL^{{\rm eff}} (B)=-\frac{1}{8\pi^2} \int_0^\infty \frac{ds}{s^3}
\left[ eBs \coth(eBs)-1-\frac13 (eBs)^2 \right] \exp(-m^2s)~~~\vspace{1ex}
\ee
 is the well known vacuum part of  the
effective Lagrangian in  the purely magnetic case
\cite{Schwinger51}.

We may rewrite the expression for the  fermion   density as
\be
	\rho =i\,  \tr (\gamma_0 G(x=x')) = 
	i  \int \frac{ d^4\!p}{(2\pi)^4}
	\tr  (\gamma_0 G(B,\mu;p))~~~,
	\label{eq-rho}
\ee
where the trace is over spinor indices only, and
$ G(x,x')$ is the fermion Green's function in configuration space.
We shall use 
 the expression obtained in \refc{ChodosEO90} for the  
Green's function in momentum space
\bea
	\lefteqn{G (B,\mu; p) =} \nonumber \\
&& - i \theta ((p_0 + \mu) {\rm sign} p_0) 
	\int_0^\infty d s\exp \left\{ i s 
	\left[ (p_0 + \mu )^2 - p_\parallel^2 -p_\perp^2 
	\frac{\tan (eBs)}{eBs} - m^2 +i
	\varepsilon
		\right]
		\right\} \times \nonumber\\
	&& \left\{ 
	\left( 1 + \gamma_1\gamma_2 \tan(eBs)			\right)
	\left[\gamma_3 p_\parallel -\gamma_0 (p_0 + \mu) - m \right] 
	+ \{ 1+ \tan^2(eBs) \} (\gamma_1 p_1 + \gamma_2p_2)
						\right\}\nonumber\\
	&& + i \theta (-(p_0 - \mu) {\rm sign} p_0) \int_0^\infty d s
	\exp 	
	\left\{- i s 
	\left[ (p_0 + \mu )^2 - p_\parallel^2 -p_\perp^2 
	\frac{\tan (eBs)}{eBs} - m^2 -i
	\varepsilon
						\right]
						\right\} \times\nonumber\\
	&& \left\{ 
	\left( 1 - \gamma_1\gamma_2 \tan(eBs)			\right)
	\left[\gamma_3 p_\parallel -\gamma_0 (p_0 + \mu) - m	\right] 
	+ \{ 1+ \tan^2(eBs) \} (\gamma_1 p_1 + \gamma_2p_2)	\right\}
%  			\nonumber\\
%	&&~~~
 \label{eq-green}
\eea
where  $p_\parallel$ and $p_\perp$ are the modulus of the momenta
parallel  and perpendicular to the magnetic field, respectively.
A different $i \varepsilon$-prescription here  arise for $|\mu| > |m|$ as 
the rules
for passing poles in the fermion Green's function are changing 
\cite{ChodosEO90,shuryak}, since 
the Dirac sea is filled up to the energy $\mu$.

Using \Eq{eq-green} in \Eq{eq-rho},  the identity 
	$\theta(x) = 1 - \theta(-x)$, and performing a
 trivial change of variables, we get
\bea
	\lefteqn{\rho (B,\mu) =
	 - \frac1{4\pi^4} \int d^4\! p \int_0^\infty d\! s \,p_0
	\exp 
	\left\{i s 
	\left[ (p_0 + \mu )^2 - p_\parallel^2 -p_\perp^2 
	\frac{\tan (eBs)}{eBs} - m^2 +i
	\varepsilon
						\right]
							\right\}} \nonumber\\
	&+&	 \frac1{4\pi^4} \int_0^\mu p_0dp_0 \int d^3\! {\bf p} 
       		\int_0^\infty ds \left( \exp 
	\left\{ i s 
	\left[ (p_0 + \mu )^2 - p_\parallel^2 -p_\perp^2 
	\frac{\tan (eBs)}{eBs} - m^2 +i
	\varepsilon
						\right]
					\right\} \right. + \nonumber \\
      &&+  \left. 
	\exp 
	\left\{- i s 
	\left[ (p_0 + \mu )^2 - p_\parallel^2 -p_\perp^2 
	\frac{\tan (eBs)}{eBs} - m^2 -i
	\varepsilon
						\right]
					\right\}\right)  ~~~,
	\label{eq-rho2}
\eea
where the first term on  the right-hand side is  vanishing. 

Performing  the momentum integration in \Eq{eq-rho2}, we get
\be
	\rho (B, \mu) = 2 \Re e
	\left\{
	\frac{e^{3i\pi/4}}{8\pi^{3/2}} \int_0^\infty \frac{ds}{s^{5/2}}
	(eBs)\cot(eBs) \exp \{i s (\mu^2 - m^2 + i \varepsilon)\}
							\right\}
\label{eq-rhochodos}
\ee
that may be  obtained from the expression for the  effective Lagrangian in 
\refc{ChodosEO90}, keeping the $i\varepsilon$ prescription. This 
$i\varepsilon$ prescription tells us that the  proper-time integral actually 
is to be performed slightly below the real axis. When closing the contour
to obtain an exponentially decreasing integrand instead of an 
oscillating one, the poles of  $\cot(eBs)$ will thus be encircled 
 when the contour is closed in the upper half-plane for
$\mu^2 > m^2$. The sum over the residues at these poles will 
form the vanishing temperature limit of the ``oscillating'' part
of the effective Lagrangian, in agreement with  \refc{ElmforsPS93}.

The proper-time integral in \Eq{eq-rhochodos}, or the corresponding integral 
after the  above-described Wick rotation, cannot be performed
analytically.
 Instead, we shall here perform the 
proper-time integral in \Eq{eq-rho2} before the integral over the momentum.
 This equation may be rewritten as~\cite{z}
\be
	\rho = \frac1{4\pi^4} \int_0^\mu p_0dp_0 \int d^3\! {\bf p} 
	\left(
	I(p) + I^*(p)
				\right)~~~,
	\label{eq-rho3}
\ee
where we have defined
\be
	I(p) = \int_0^\infty ds \exp 
	\left\{ 
	i s 
	\left[
	p_0^2 - p_\parallel^2 - p_\perp^2 \frac{\tan(eBs)}{eBs} - m^2 +i
	\varepsilon
						\right]\right\}~~~.
	\label{eq-i}	
\ee
For $m^2 > p_0^2 - p^2_\parallel$ we may close the integration contour in the
lower half plane
\be
	I(p) = - i \int_0^\infty ds \exp 
	\left\{ 
	- s 
	\left[
	m^2 - p_0^2 + p_\parallel^2 + p_\perp^2 \frac{\tanh(eBs)}{eBs} -i
	\varepsilon
						\right]\right\}~~~.
	\label{eq-ii}	
\ee

The integral in \Eq{eq-ii} is
diverging  as $s \rightarrow \infty$ for $m^2 \le p_0^2 - p_\parallel^2$. 
Changing  the variable of integration to  $ z = {\rm tanh} (eBs) \quad
 (eB >0)$, and defining 
$Q = (m^2 + p_\parallel^2 -p_0^2 - i \varepsilon)/2eB$,
 we may rewrite \Eq{eq-ii} as  \cite{z1}
\be
	I(p) = - \frac{i}{eB}  \int_0^1 dz (1 - z)^{-1 +Q} (1+ z)^{-1-Q}
	\exp \left\{ - \frac{p_\perp^2}{eB} z			\right\}~~~, 
	\label{eq-iii}	
\ee
that has a  singularity at $z=1$.
 From the left- and  right-hand sides of  \Eq{eq-iii} one may now extract
the first $k$
terms of the Taylor expansion of  
$(1+z)^{-1-Q}\exp\{-p_\perp^2z/eB\}$ around  $z=1$ (Cauchy method),
\bea
	\lefteqn{I(p) + \frac{i}{eB} \sum_{n=0}^k 
	\int_0^1dz a_n(p) (1-z)^{n+Q-1} = } \nonumber \\
	&& -\frac{i}{eB} \left\{
	(1+ z)^{-1-Q} 
	\exp \left\{ - \frac{p_\perp^2}{eB} z			\right\}
	- \sum_{n=0}^k 
	\int_0^1dz a_n(p) (1-z)^n 				\right\}
	(1-z)^{-1+Q}~~~,
	\label{eq-iiii}	
\eea
where we have defined
\be
   a_n(p)\equiv \frac{(-1)^n}{n!} \, \frac{d^n}{dz^n} \left. \left[
    (1+z)^{-1-Q}\exp\{-p_\perp^2z/eB\} \right] \right|_{z=1}~~~.
\ee
Performing the  trivial integration in the left-hand side of \Eq{eq-iiii},
under the assumption that $Q + n > 0$, we get
\bea
	\lefteqn{I(p) + \frac{i}{eB} \sum_{n=0}^k 
	\frac{a_n(p)}{n +Q} (1-z)^{n+Q-1} =}  \nonumber\\
	&&  - \frac{i}{eB}  \int_0^1 dz 
	\left\{
	(1+ z)^{-1-Q} 
	\exp \left\{ - \frac{p_\perp^2}{eB} z			\right\}
	- \sum_{n=0}^k 
	\int_0^1dz a_n(p) (1-z)^n 				\right\}
	(1-z)^{-1+Q}~~~.\nonumber \\
	\label{eq-i5}	
\eea
The integral in the right-hand side of \Eq{eq-i5} is  convergent in  the
halfplane  $\Re e(Q) > $ \mbox{$-(k+1)$}. 
Taking limit $k \rightarrow \infty$ we see that \Eq{eq-i5}
 is an analytical continuation of $I(p)$ on the whole complex plane
 $p_0^2 - p_\parallel^2$ , excluding the  points  
$p_0^2 - p_\parallel^2 = m^2 + 2eBn,\quad
 n =
0,1,2, \dots $,\quad where $I(p)$ has simple poles
(these poles are nothing but the familiar 
relativistic Landau levels). 

	Substituting the expression for $I(p)$ rewritten as in \Eq{eq-i5} into
\Eq{eq-rho3} we  see that the regular parts of $I(p)$ and  $I^*(p)$ cancel,
and thus
\bea
	\lefteqn{ \rho (B, \mu) =
	 \frac{i}{2\pi^4} \int_0^\mu p_0dp_0 \int d^3{\bf p}
	\sum_{n=0}^\infty \times }  \nonumber\\
	&& \left(
	\frac{a_n(p)}{p_0^2 - p_\parallel^2 - m^2 - 2eBn +i \varepsilon} -
	\frac{a_n^*(p)}{p_0^2 - p_\parallel^2 - m^2 - 2eBn -i \varepsilon} 
							\right)~~~.
\eea
Using the  identity $(x \pm i \varepsilon)^{-1} = \wp (x^{-1} ) \mp i \pi
\delta(x)$, we see that also the principal values cancel, 
and the only nonvanishing contribution comes from the poles
\bea
	\rho (B, \mu)& =& \frac{1}{\pi^3} \int_0^\mu p_0dp_0 \int d^3{\bf p}
	\sum_{n=0}^\infty 
	a_n(p)\delta(p_0^2 - p_\parallel^2 - m^2 - 2eBn)  \nonumber\\
	&=& \frac{1}{2\pi^2} \int_{-\infty}^{\infty}dp_\parallel
	\int_0^\infty d (p^2_\perp)
	\sum_{n=0}^\infty 
	a_n(p_{(n)})\theta(\mu^2 - p_\parallel^2 - m^2 - 2eBn)~~~,
\eea
where $p_{(n)}$ denotes the four-momentum, such that $(p_{0(n)})^2=m^2 +  (p_{\parallel(n)})^2 +2eB n$. 
The  Heavyside step-functions here  does accordingly
describe the number of (partially) filled Landau levels.
For example, for $0 < \mu^2 - m^2 <2eB$, only the lowest Landau level ($n = 0$)
  contributes to the density
\bea
	\rho_0 (B, \mu)& =&
		 \frac{1}{2\pi^2} \int_{-\infty}^{\infty}dp_\parallel
	\int_0^\infty d (p^2_\perp)
	a_0(p_0)~\theta(\mu^2 - p_\parallel^2 - m^2 - 2eBn)  \nonumber\\
%	&=& \frac1{\pi^2} \sqrt{\mu^2 - m^2}~\theta(\mu^2 - m^2) \frac12 
%	 \int_0^\infty d (p^2_\perp) \exp 
%	\left( 
%	-\frac{p^2_\perp}{eB} \right) \nonumber \\
	 &=& \frac{eB}{2\pi^2}
	 \sqrt{\mu^2 - m^2}~\theta(\mu^2 - m^2) ~~~.
	\label{d0}
\eea
For $0 < \mu^2 - m^2 <4eB$, we get similarly 
 $\rho(B,\mu) = \rho_0(B,\mu) + \rho_1(B,\mu)$, where
\bea
	\rho_1 (B, \mu) &=&
	\frac1{\pi^2} \sqrt{\mu^2 - m^2 - 2eB}~\theta(\mu^2 - m^2 - 2eB)  
	 \int_0^\infty d (p^2_\perp)  \frac{p^2_\perp}{eB}  \exp 
	\left( 
	-\frac{p^2_\perp}{eB}
						\right)\nonumber\\
	& =& \frac{eB}{\pi^2}
	 \sqrt{\mu^2 - m^2 - 2eB}~\theta(\mu^2 - m^2 - 2eB)~~~.
	\label{eq-d1}
\eea
The  general expression for the fermion density in a static 
uniform  magnetic field $B$, may thus be written as a sum over a finite
number  of occupied Landau levels
\bea
	\rho(B,\mu) = 	\sum_{n=0}^{\left[  \frac{\mu^2 - m^2}{2eB}
								\right]}
        \rho_n(B,\mu)~~~,
	\label{eq-density}
\eea
where $[ \dots ]$ denotes the integral part. The contribution from the $n$-th
Landau level is
\be
	\rho_n(B,\mu) = b_n \, \frac{eB}{2\pi^2} 
	 \sqrt{\mu^2 - m^2-2eBn } ~~~, 
\label{eq-dens}
\ee
where we have defined
\be
  b_n\equiv 2- \delta_{n,0}~~~,
\ee
since the lowest Landau level ($n=0$), unlike the higher levels,
only contains fermions with one projection of the spin; as found in the
above examples of $\rho_0$ and $\rho_1$, cf. \Eq{eq-landenergy}.

We have thus  found an expression for the fermion density in a nonvanishing 
average magnetic field,
in terms of elementary functions in a discrete finite 
 sum over filled Landau levels. This result may be well understood from the
index theorem approach \cite{niemi}. The  density (fermion number) 
depends on the
difference of numbers of filled  positive and negative energy levels, 
which in this case is
the (semidescrete) number of Landau levels in the interval $[0,\mu]$.

	In  the limit of vanishing  magnetic field, $eB \rightarrow 0$, the
Riemann sum of  \Eq{eq-dens} may be rewritten as an integral
\be
	\rho(\mu) = \frac{1}{2\pi^2} \int_0^{\mu^2 - m^2} dx
	 (\mu^2 - m^2 - x)^{1/2} \theta (\mu^2 - m^2)= \frac1{3\pi^2} 
	\left(
	\mu^2 - m^2			\right)^{3/2} \theta (\mu^2 - m^2)  
\ee 
which is the  familiar expresion for the fermion density.

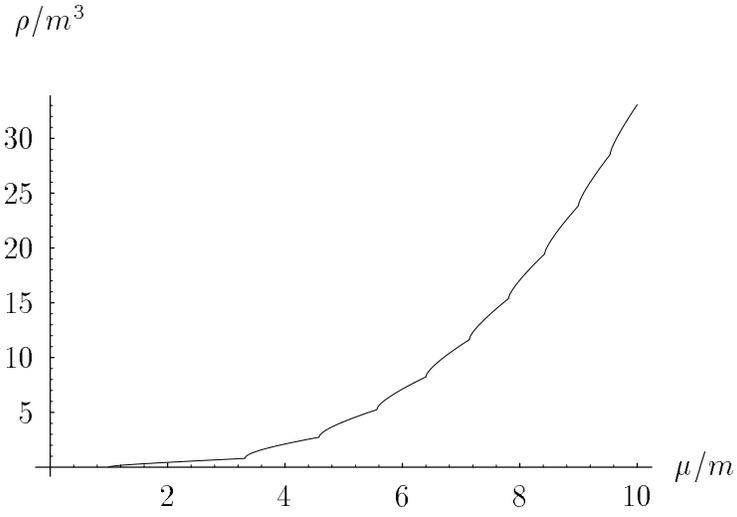
In Figure 1 we show the density as a function of the 
chemical potential for fixed magnetic field, and in Figure 2
the density is given as a function of the magnetic field for fixed
 chemical potential.  We see that the density is showing an oscillating
 behaviour as consecutive Landau levels are passing the Fermi level.

Integrating \Eq{eq-density} with respect to the chemical potential, we find
the part of the effective Lagrangian  due to the finite density as
\be
  \lm=\sum_{n=0}^{\left[  \frac{\mu^2 - m^2}{2eB} \right]}
        \cL_n(B,\mu)~~~,
\label{eq-mulag}
\ee
where the  contribution from the $n$-th Landau level is
\be
   \cL_n(B,\mu)= b_n \, \frac{eB}{4\pi^2} \left\{ \mu \sqrt{\mu^2-m^2-2eBn}
    - (m^2+2eBn) \ln \left( \frac{ \mu + \sqrt{\mu^2-m^2-2eBn}}
    {\sqrt{m^2+2eBn}}\right) \right\}~~~.
\ee

Here we have used the zero temperature proper-time method to calculate the
fermion density and  effective Lagrangian. It is noteworthy to 
compare with the approach of quantum statistical mechanics.
 By comparing the generating 
functional for fermionic Green's functions in imaginary time \cite{Fradkin59},
with the partition function in the grand canonical ensemble ($Z$), we find that
$\tilde\cL^{\rm eff}=\frac{1}{\beta V} \ln Z$. The relativistic 
fermion energy-levels in a 
static  uniform magnetic field  are found as \cite{ItzyksonZ}
\be
  E_{k,\lambda}(p_\parallel)= \sqrt{m^2 +p_\parallel^2 + 2eB(k+\lambda-1)}~~~,
\label{eq-landenergy}
\ee
where $k=0,1,2, \ldots$ corresponds to the quantized 
 orbital angular momentum, and
$\lambda=1,2$ describes the projection of  spin. Using the ordinary 
relativistic dispersion law to reintroduce the momenta orhogonal to the 
magnetic field, we find the density of states $V \, eB/(2\pi)^2 $, and readily 
obtain
\bea
	\lefteqn{\tilde\cL^{\rm eff}(B,\mu,T)=} \nonumber \\
 && \frac1{\beta}\frac{eB}{(2\pi)^2} \sum_{k=0}^\infty
\sum_{\lambda=1}^2 \int_{-\infty}^\infty d\!p_\parallel \left\{
    \ln[ 1+e^{-\beta(E_{k,\lambda}(p_\parallel)-\mu)}] +
      \ln[ 1+e^{-\beta(E_{k,\lambda}(p_\parallel)+\mu)}] \right\}~~~.
\label{eq-fieldpart}
\eea
Integrating by parts with respect to  $p_\parallel$ 
in \Eq{eq-fieldpart}  we find
\be
  \tilde\cL^{\rm eff}(B,\mu,T) =  \frac{eB}{(2\pi)^2} \sum_{k=0}^\infty
\sum_{\lambda=1}^2 \int_{-\infty}^\infty dp_\parallel \frac{p_\parallel^2}
{ E_{k,\lambda}(p_\parallel)} \left\{
    \frac1{1+e^{\beta(E_{k,\lambda}(p_\parallel)-\mu)}} +
      \frac1{ 1+e^{\beta(E_{k,\lambda}(p_\parallel)+\mu)}} \right\}~~~.
\ee
Using 
$1/(1+e^{\beta(E_{k,\lambda}(p_\parallel)\mp \mu)}) \rightarrow 
\theta[\pm \mu -  E_{k,\lambda}(p_\parallel)] $, as $\beta \rightarrow \infty$,
we may in the limit of  vanishing temperature  perform the momentum integral  
and  arrive at \Eq{eq-mulag}.\bigskip

The magnetisation of the fermion gas is now easily found
 by performing the derivative with respect to the
magnetic  field, $\tilde M=\frac{\partial}{\partial B}\tilde\cL^{\rm eff}$,
with the result
\be
  \tilde M(B,\mu)=\sum_{n=0}^{\left[  \frac{\mu^2 - m^2}{2eB} \right]}
        M_n(B,\mu)~~~,
\label{eq-magn}
\ee
where the  contribution from the $n$-th Landau
  level is
\be 
  M_n(B,\mu)=  b_n \, \frac{e}{4\pi^2} \left\{ \mu  \sqrt{\mu^2-m^2-2eBn}
		- (m^2+4eBn) \ln \left( \frac{ \mu + \sqrt{\mu^2-m^2-2eBn}}
    {\sqrt{m^2+2eBn}}\right)  \right\}~~~.
\ee
We notice that $\lim_{(\mu^2-m^2-2eBn \rightarrow 0^+)}\! \cL_n(B,\mu)=0$, and
 $\lim_{(\mu^2-m^2-2eBn \rightarrow 0^+)}\! M_n(B,\mu)=\nolinebreak0$, so
that the Lagrangian density as well as the  magnetisation are continuous.
Figure 3 shows the total
magnetisation, $M^{\rm tot}(B,\mu) \equiv \frac{\partial}{\partial B}
 \cL^{\rm eff}(B,\mu)$ \cite{ElmforsPS93},
 as a function of the magnetic field for fixed chemical
 potential (however, the vacuum contribution $
\frac{\partial}{\partial B}\cL^{\rm eff}(B)$ is small in this range of
 parameters).
In agreement with \refs{Canuto68,ElmforsPS93,LeeCCC69}, we see that for
low temperatures (here $T=0$), the relativistic fermion gas exhibits the 
de Haas--van Alphen effect.\bigskip 

\section*{Acknowledgement}
D.P. wants to thank Per Elmfors, Per Liljenberg, Per Salomonsson and 
 Bo-Sture Skagerstam for fruitful discussions. V.Z. 
is grateful to Sergey Rashkeev and Alexander Zagoskin for illuminating
 discussions,
 and Prof. Lars Brink 
for his kind hospitality at the Institute of Theoretical Physics, 
Chalmers University of Technology and  G\"oteborg  University, where this
 work was done. V.Z. 's  work was 
supported in part by Soros Humanitarian Foundation 
Grant awarded by the  American Physical Society and Russian Fund of 
Fundamental Researches
 Grant $N^o$ 67123016.

\bigskip
%\bibliography{}

\begin{thebibliography}{99}
%
\bibitem{Chanm92}
     G.~Chanmugam,
    {\it Ann.~Rev. Astron.~Astrophys.}~{\bf 30} (1992) 143.
%
\bibitem{ShapiroT83}
      S. L. Shapiro and S. A. Teukolsky, ``{\it Black Holes, White
      Dwarfs and Neutron Stars, The Physics of Compact Objects}'', 
      (Wiley, New York, 1983).
%
\bibitem{Canuto68}
     V.~Canuto and H.-Y.~Chiu, \prl{21}{1968}{110};\\
  V.~Canuto and H.-Y.~Chiu,
\pr{173}{1968}{1210,1220,1229};\\
V.~Canuto, H.-Y.~Chiu and L.~Fassio-Canuto,		      
\pr{176}{1968}{1438}.
%
%
\bibitem{Cabo81}
    A.~Cabo, 
\fort{29}{1981}{495}
%
\bibitem{ChodosEO90}
      A. Chodos, K. Everding and D. A. Owen,
      \pr{D42}{1990}{2881}.
%
\bibitem{ElmforsPS93}
      P.~Elmfors, D.~Persson and
      B.-S.~Skagerstam,  \prl{71}{1993}{480} ;\\
       P.~Elmfors, D.~Persson and
      B.-S.~Skagerstam, preprint ( NORDITA-93/78 P, G\"oteborg ITP 93-11, 
hep-ph/9312226).
%
\bibitem{LeeCCC69}
  H.~J.~Lee,  V.~Canuto, H.-Y.~Chiu and C.~Chiuderi, \prl{23}{1969}{390}.
%
%
\bibitem{Schwinger51}
      J. Schwinger, \pr{82}{1951}{664}.
%
%
\bibitem{shuryak} V.~E.~Shuryak, \prep{61}{1980}{71}. 

%
\bibitem{z}
	Vad.~Y.~Zeitlin, \mpl{A8}{1993}{1821}.
%%
\bibitem{z1}
	Vad.~Y.~Zeitlin, {\it Sov.~J.~Nucl.~Phys.}\/[Yadernaya Fizika]  
{\bf 49} (1989) 742.
%
%
\bibitem{niemi}
A.~J.~Niemi, \np{B251}{1985}{155}.
%
\bibitem{Fradkin59} E.~S.~Fradkin, \np{12}{1959}{465}.
%
\bibitem{ItzyksonZ}
C.~Itzykson and J.-B.~Zuber, ``{\it Quantum Field Theory}'', 
(McGraw-Hill,1980).
%

\end{thebibliography}

%
\end{document}